\documentclass[12pt,prl,superscriptaddress,showpacs,preprintnumbers,
amsmath,amssymb]{revtex4}
\usepackage{graphicx} % Include figure files
\usepackage{dcolumn} % Align table columns on decimal point
\def\b0{{\bf 0}}

\begin{document}

\title{Capillary-wave models and the effective average action scheme of functional renormalization group}

\author{P.~Jakubczyk}
\email{pawel.jakubczyk@fuw.edu.pl}
\affiliation{Institute of Theoretical Physics, Faculty of Physics, University of Warsaw, 
 Ho\.za 69, 00-681 Warsaw, Poland}

\date{\today}

\begin{abstract}
We reexamine the functional renormalization-group theory of wetting transitions. As a starting point of the analysis we apply an exact  
equation describing renormalization group flow of the generating functional for irreducible vertex functions. We show how the standard nonlinear
renormalization group theory of wetting transitions can be recovered by a very simple truncation of the exact flow equation. The derivation makes 
all the involved approximations transparent and demonstrates the applicability
of the approach in any spatial dimension $d\geq 2$. Exploiting the
non-uniqueness of the renormalization-group cutoff scheme, we find however, that the capillary parameter $\omega$ is a scheme-dependent quantity below $d=3$. For $d=3$ the parameter $\omega$ is perfectly robust against scheme variation. 

\end{abstract}
\pacs{05.10.Cc, 68.03.-g, 68.08.Bc}

\maketitle

\section{Introduction}
Wetting transitions \cite{Cahn_77, Ebner_77} have posed a considerable challenge to condensed matter theory over the last decades. Of particular 
focus is the controversial critical wetting transition \cite{Parry_09} in systems with short-ranged intermolecular interactions. This is of remarkable interest 
to the renormalization-group (RG) theory. The physical dimensionality $d=3$ turns out to be the upper critical dimension 
yielding subtle, non-universal behavior very distinct to bulk criticality. The transitions in question are specific in a number of respects: (1) critical exponents are non-universal; (2) fluctuation effects are 
important in $d\in [2,3]$ even though anomalous scaling of the corresponding propagator is absent (the critical exponent $\eta=0$); (3) even though 
fluctuation effects are important, they give rise to no phase boundary shifts in many physical cases in $d$ close to or equal three; (4) in the 
so-called strong-fluctuation regime, the interfacial correlation length diverges exponentially at the transition in $d=3$ (at the upper critical dimension!), while for $d\in [2,3[$ and above $d=3$ it displays a power-law singularity. 

Renormalization group theory has proven an efficient tool to analyze the remarkable fluctuation effects near the wetting transitions. The early 
studies \cite{Brezin_83, Fisher_85} were performed in different formulations and included only terms linear in the effective interfacial 
potential. Although such an approach ultimately yields correct critical behavior in $d=3$, it fails to controllably describe the strongly repelling 
interaction between the interface dividing the coexisting phases and the solid wall at short separations. For $d<3$ nontrivial fixed points 
describing the disordered phase and the critical manifold in the functional space are missed by a linearized RG framework. So are fixed points 
describing multicritical behavior.

A significant improvement was offered in Refs.~\cite{Lipowsky_86, Lipowsky_87}, where a non-linear RG procedure was carried out relying on 
Wilson's approximate recursion relations. The study yields a general, unified description of a wide class of models featuring wetting-type 
transitions. In particular, it shows that the transitions in the strong-fluctuation regime, occurring for quite distinct interfacial models, are described by just one fixed point. The study was subsequently extended to account for multicritical phenomena \cite{David_90}.
This theory relies on approximate recursion relations, derived following Wilson's scheme making a number of  approximations. Below we rederive the continuous version of these relations using the one-particle irreducible variant of 
functional RG. The derivation is very simple and relies on clear and modest
assumptions. We subsequently exploit the ambiguity in defining the RG
scheme. We analyze the linear RG limit and argue that the capillary parameter
$\omega$, influencing critical properties and determining the boundaries between different scaling regimes is a scheme-dependent quantity below $d=3$. Subsequently we show that no such dependence occurs at the physical dimensionality $d=3$.

\section{The exact RG flow equation}
In the context of critical wetting phenomena, we consider a capillary-wave type effective Hamiltonian
\begin{equation}
\label{Hamiltonian}
 \mathcal{H}[ \tilde{l}]=H[ \tilde{l}]/k_BT=\int d^{d-1}x\left[\frac{\sigma}{2}\left(\nabla \tilde{l}\right)^2+
V(\tilde{l})\right]\;,
\end{equation}
where $\tilde{l}(\bf{x})$ describes the position of a fluctuating interface dividing two coexisting phases above a point $\bf{x}$ on a planar, 
uniform wall. The interfacial stiffness $\sigma$ controls the energetic cost of deforming the interface with respect to a planar configuration. 
The effective interfacial potential $V(\tilde{l})$ is infinite or very large for $\tilde{l}<0$ to prevent the interface from 
penetrating the wall. It displays a minimum at $\tilde{l}=\bar{l}$ corresponding to the mean-field (MF) value of the interfacial separation, and approaches a constant value as $\tilde{l}\to \infty$. 
By varying temperature or another control parameter the system may be tuned to a wetting transition, where the average value of $\tilde{l}$ 
continuously diverges. In the MF and weak-fluctuation scaling regimes (see \cite{Fisher_85, Lipowsky_87}) this coincides with the vanishing of 
the attractive tail in $V(\tilde l)$. The capillary-wave models are capable of describing only large-scale interfacial physics. The  
variables should be understood as coarse-grained over a scale of the order of the intrinsic width of the interface, which in turn is of the 
order of the bulk correlation lengths. The Hamiltonian Eq.~(\ref{Hamiltonian}) is therefore supplemented by an upper momentum cutoff $\Lambda_0$.        

The theoretical framework we employ is usually referred to as one-particle irreducible functional RG, or effective action RG. For the 
remaining part of the section we outline a derivation of an exact flow equation, following Refs.~\cite{Wetterich_93, Berges_02}.

We first consider the functional 
\begin{equation}
 W_\Lambda\left[h\right]=\log\int\mathcal{D}\tilde{l}e^{-\mathcal{H}[\tilde{l}]-\Delta \mathcal{H}_\Lambda[\tilde{l}]+\int d^{d-1}xh(\mathbf{x})  \tilde{l}
 (\mathbf{x})}\;.
\end{equation}
When $\Delta \mathcal{H}_\Lambda[\tilde{l}]$ vanishes, $W_\Lambda\left[h\right]$ is just the generating functional for the connected correlation functions.
The additional term $\Delta \mathcal{H}_\Lambda[\tilde{l}]$ is assumed quadratic in $\tilde{l}$
\begin{equation}
\Delta \mathcal{H}_\Lambda [\tilde{l} ]=\frac{1}{2}\int d^{d-1}x\int d^{d-1}y R_\Lambda (x,y)\tilde{l}(\mathbf{x})\tilde{l}
 (\mathbf{y} )=\frac{1}{2}\int\frac{d^{d-1}q}{
(2\pi)^{d-1}}R_\Lambda (q)\tilde{l}_{\mathbf{q}}\tilde{l}_{-\mathbf{q}}\;. 
\end{equation}
The cutoff function $R_\Lambda (q)$ is required to vanish when the cutoff scale $\Lambda$ is sent to zero, and become infinite (or very large) for 
$\Lambda\to \Lambda_0$ at any fixed wavevector $\bf{q}$. Due to the presence of the term $\Delta \mathcal{H}_\Lambda$, the Fourier modes of $\tilde{l}$ corresponding 
to sufficiently small $q$ ($q<\Lambda$) aquire an effective $q$-dependent mass providing an infra-red cutoff. The short modes ($q>\Lambda$) are 
not affected by the term $\Delta \mathcal{H}_\Lambda$. 

The scale-dependent effective action is defined via the modified Legendre transform:
\begin{equation}
\label{ef_act}
\Gamma_\Lambda [l(\mathbf{x})]=-W_\Lambda[h(\mathbf{x})]+\int d^{d-1} x h(\mathbf{x}) l(\mathbf{x})-\Delta \mathcal{H}_\Lambda [l(\mathbf{x})]\;, 
\end{equation}
 where
\begin{equation}
 l(\mathbf{x})=\langle \tilde l (\mathbf{x}) \rangle = \frac{\delta W_\Lambda [h]}{\delta h (\mathbf{x})}\;. 
\end{equation}
The last term in Eq.~(\ref{ef_act}) provides that $\Gamma_{\Lambda_0} [l]=\mathcal{H}[l]$ (see e.g. Ref.~\cite{Berges_02}). 
The quantity $\Gamma_\Lambda [l]$ interpolates between the Hamiltonian (for $\Lambda=\Lambda_0$) and the usual Legendre transform of the 
logarithm of the partition function, i.e. the Gibbs free energy for $\Lambda\to 0$, when the infrared cutoff is removed. 
Executing the derivative of Eq.~(\ref{ef_act}) with respect to $\Lambda$, after simple algebra we obtain the following flow equation \cite{Wetterich_93}
\begin{equation}
\label{exact_flow_eq}
\partial_{\Lambda}\Gamma_\Lambda [l]=\frac{1}{2}\int\frac{d^{d-1}q}{(2\pi)^{d-1}}\frac{\partial_\Lambda R_\Lambda (q)}{\Gamma_\Lambda^{(2)}[l]+
R_\Lambda(q)}\;,
\end{equation}
where 
\begin{equation}
 \Gamma_\Lambda^{(2)}[l]=\frac{\delta^2 \Gamma_\Lambda [l]}{\delta l_{\mathbf{q}} \delta l_{-\mathbf{q}} }\;.
\end{equation}
Solution of Eq.~(\ref{exact_flow_eq}) with the Hamiltonian $\mathcal{H}[l]$ as the initial condition is equivalent to evaluating the corresponding partition function. 
  
\section{Truncation}
The functional differential equation Eq.~(\ref{exact_flow_eq}) can hardly ever be solved exactly. However, in many contexts it provides an interesting starting 
point for non-standard approximation schemes that may go beyond perturbation theory. Below we show how the basic equation of the non-linear theory 
of wetting is simply recovered from Eq.~(\ref{exact_flow_eq}) relying on two assumptions: (1) locality of the 
effective action, (2) neglecting the renormalization of 
the stiffness coefficient $\sigma$ and higher order gradient terms. The discussion of these assumptions is postponed to the last section. 

Relying on the assumed locality, we expand the effective action in gradients of $l(\mathbf{x})$. 
\begin{equation}
\label{deriv_exp}
\Gamma_\Lambda [l]=\int d^{d-1}x\left[U_\Lambda (l)+\frac{1}{2}\Sigma_\Lambda(l)(\nabla l)^2+...\right]\;, 
\end{equation}
and subsequently neglect the renormalization of the gradient terms, putting $\Sigma_\Lambda (l)=\sigma$ for all 
$\Lambda$. Fully analogous approach was developed for bulk criticality described by the $O(N)$ models  \cite{Berges_02, Ballhausen_04, Canet_03}. In that context the order parameter field shows an anomalous dimension at criticality (the exponent $\eta\neq 0$) for $d<4$ and therefore the renormalization of the gradient term should not 
be neglected. Accurate calculation of the critical exponents with this method in $d=3$ requires accounting for terms of the order $\partial^4$. 
The present situation is completely different, as $\eta=0$ for any $d$ at the wetting 
transition \cite{Lipowsky_87}. What is neglected is therefore only a finite renormalization of the stiffness 
coefficient. The approach should therefore yield very accurate results. 

By plugging the derivative expansion Eq.~(\ref{deriv_exp}) into Eq.~(\ref{exact_flow_eq}) we arrive at a flow equation 
for the effective potential $U_\Lambda (l)$:
\begin{equation}
\label{U_flow}
\partial_\Lambda U_\Lambda (l) =\frac{1}{2}\int\frac{d^{d-1}q}{(2\pi)^{d-1}}\frac{\partial_\Lambda R_\Lambda (q)}{\sigma q^2+U_\Lambda ''(l)+R_\Lambda(q)}\;,  
\end{equation}
where 
\begin{equation}
 U_\Lambda ''(l) =\frac{\partial^2 U_\Lambda (l)}{\partial l^2}\;.
\end{equation}
We now introduce the following variables:
\begin{equation}
s=-\log\frac{\Lambda}{\Lambda_0}\;,\;\;\; y=\frac{q^2}{\Lambda^2}\;,\;\;\; z=\sqrt{\sigma}\Lambda^{\frac{3-d}{2}} l\;,\;\;\; u_\Lambda (z)=U_\Lambda (l)\Lambda^{-d+1}\;,\;\;\; r_\Lambda (y)=\frac{R_\Lambda(q)}{\sigma q^2}\;,
\end{equation}
in terms of which Eq.~(\ref{U_flow}) takes the scale-invariant form:
\begin{equation}
\label{Int_eq}
\partial_s u_\Lambda (z)=\frac{S_{d-2}}{4(2\pi)^{d-1}}\int dy y^{\frac{d-3}{2}}\frac{\partial_s r_\Lambda (q)}{1+y^{-1}
\partial^2_z u_\Lambda (z)+r_\Lambda (y)} + (d-1)u_\Lambda (z) + \frac{3-d}{2} z\partial_z u_\Lambda (z)\;,
\end{equation}
where we performed integration over the angular variables, and $S_d=\frac{2
  \pi^{\frac{d+1}{2}}}{\Gamma (\frac{d+1}{2})}$ is the surface area of a
$d$-dimensional sphere of radius one. We stress, that Eq.~(\ref{Int_eq}) was
derived assuming not more than that $\sigma$ remains constant during the flow
and the local structure of the effective action is preserved by the flow. The
choice of the cutoff function is to a large extent arbitrary
\cite{Berges_02}. As long as no approximations are made, the results obtained
in the limit $\Lambda\to 0$ do not depend on the particular form of $R_\Lambda
(q)$. Truncation of the exact flow equation introduces a weak cutoff
dependence of the computed quantities, which can serve the purpose of
estimating the errorbars.  

We now specify the limit, in which the standard theory is recovered. We consider the sharp momentum cutoff, where $r_\Lambda (y)\to 0$ for $y>1$ and $r_\Lambda (y)\to \infty$ for
$y<1$. This is realized by taking
\begin{equation}
\label{sharp_cutoff} 
R_\Lambda (q)=\lim_{\gamma\to\infty}\sigma\Lambda^2\gamma\theta (\Lambda^2-q^2)\;.
\end{equation}
With this choice, the remaining integral in Eq.~(\ref{Int_eq}) can be done. This yields
\begin{equation}
\label{Lipowsky_eq}
\partial_s u_\Lambda(z)=\frac{S_{d-2}}{2(2\pi)^{d-1}}\log (1+\partial_z^2 u_\Lambda(z)) + (d-1)u_\Lambda(z) +\frac{3-d}{2} z\partial_z u_\Lambda (z)+const\;,
\end{equation}
where the constant term shifts the flowing potential $u_\Lambda(z)$ without influencing its shape and can therefore be skipped.
After making the following substitution of variables:
\begin{equation}
u_\Lambda=\frac{v}{k_B T\Lambda_0^{d-1}}\; ,\;\;\;\;\; z=\sqrt\frac{\sigma}{k_B T\Lambda_0^{d-3}}z'\;,
\end{equation}
we obtain
\begin{equation}
\partial_s v(z')=\frac{\Lambda_0^{d-1}k_BT}{(4\pi)^{(d-1)/2}\Gamma(\frac{d-1}{2})}\log (1+\frac{1}{\sigma\Lambda_0^2}\partial_{z'}^2v(z')) + 
(d-1)v(z') +\frac{3-d}{2} z'\partial_{z'} v(z')\;,
\end{equation}
which coincides with the continuous limit of the Wilson's approximate recursion relations as derived and analyzed for wetting 
phenomena in Ref.~\cite{Lipowsky_87}.
After expanding the logarithm we recover the linear flow equation analyzed in Ref.~\cite{Fisher_85}. 

Although the present analysis yields the same result as Ref.~\cite{Lipowsky_87} it avoids approximations made while performing the partial trace 
over the short modes in the Wilson scheme. The ``arbitrary'' length scale, which in \cite{Lipowsky_87} is determined using phenomenological arguments and so that the 
linear RG \cite{Fisher_85} is recovered in the appropriate limit is absent here. We evaded the need for any approximations regarding the eigenfunctions 
into which the short modes are expanded. In the present framework, the basic nonlinear RG equation of wetting phenomena turns out to emerge 
in the sharp-cutoff limit from an exact flow equation merely by demanding that the stiffness coefficient is not renormalized. 
Also note, that the linear RG equation \cite{Fisher_85} is recovered by neglecting terms only of the order $(\partial_{z'}^2v(z'))^2$. In fact, the order of these terms does not depend on the choice of the cutoff. This yields the conclusion, that the reliability of the linear RG approach does not depend on the magnitude of $V(\tilde{l})$ and $\partial_{\tilde{l}}V(\tilde{l})$ in the vicinity of the wall, and the neglected terms are only quadratic in curvature.

\section{Cutoff scheme dependencies}
We now pass to another point of this paper. It has to be emphasized, that
despite the modesty of the approximations made, the equation defining the
non-linear theory of wetting cannot be regarded as unique. Indeed, the
Lipowsky-Fisher framework is recovered only in the sharp cutoff limit given by
Eq.~(\ref{sharp_cutoff}). Exploiting the freedom of the choice of the cutoff
function $R_\Lambda (q)$, a whole family of flow equations can be
generated. Variation of the scale parameter $\Lambda$ corresponds to integrating out a different set of modes depending on the cutoff $R_\Lambda (q)$. The fact that different regulators yield distinct flow equations is not particularly surprising since for two distinct choices of $R_\Lambda (q)$ the corresponding effective potentials considered at a scale $\Lambda$ describe two different physical systems. The present formulation of functional RG assures that the RG trajectories are universal in the so called freezing regime, once the cutoff scale is lowered below the scale set by renormalized masses present in the problem and the tree contributions to the beta-functions dominate. For a detailed discussion as well as connection to other exact RG frameworks (Wilson-Polchinski and Wegner-Houghton) see Ref.~\cite{Sumi_99}. We note however that the 3d wetting problem is special in that the critical behavior is non-universal. The critical exponents depend on the capillary parameter $\omega$ \cite{Fisher_85}, which in turn involves the dependence on microscopic details of the system. In consequence, these cannot be viewed as completely irrelevant for the problem. 

One very convenient and widely applied cutoff choice, often referred to
as Litim cutoff \cite{Litim} is
\begin{equation}
R_\Lambda (q) =\sigma (\Lambda^2-q^2)\theta(\Lambda^2-q^2) \; ,
\end{equation} 
which leads to the following flow equation
\begin{equation}
\label{Litim_fl_eq}
\partial_s u_\Lambda (z) = -\frac{A_d}{1+\partial^2_z u_\Lambda (z)} + (d-1)u_\Lambda (z) + \frac{3-d}{2}z\partial_z u_\Lambda (z) \; ,
\end{equation}
where $A_d=\frac{S_{d-2}}{(d-1)(2\pi)^{d-1}}$. This is not particularly surprising, that
the non-linear part of the flow equation differs as compared to
Eq.~(\ref{Lipowsky_eq}). More importantly however, upon expanding in
$\partial_z^2 u_\Lambda$ the linear RG is again recovered, but the coefficient in
front of the $\partial^2_z u_\Lambda (z)$ term is now different by the factor
$B=\frac{2}{d-1}$, yielding the conclusion that the capillary parameter
$\omega\sim B^{-1}$ is a scheme-dependent quantity. 

This observation implies that the linearized RG framework involves a
dependence of the crucial parameter of the theory on the approximation
scheme. In the physically most relevant case $d=3$ one finds $B=1$ and
consequently the two cutoff choices 
discussed above yield the same value of the capillary parameter. This is quite
unlike the case $d=2$, where the linear RG is however not an adequate
approach. The present observation would give rise to sensitivity of critical
behavior to the cutoff scheme for $d\in ]2,3[$ when these are computed within
the linearized theory. This dependence ceases as $d\to 3^-$. For $d>3$ the
critical exponents carry no dependence on $\omega$ are are determined by
mean-field theory.  

The robustness of the capillary parameter in $d=3$ against scheme variation can in fact be proven witout specifying the form of the cutoff function $R_\Lambda (q)$. To this aim we inspect Eq.~(\ref{Int_eq}) and perform a linearization of the integrand
\begin{equation}
\label{Expanded}
 \partial_s u_\Lambda (z)=\frac{S_{d-2}}{4(2\pi)^{d-1}}\int dy y^{\frac{d-3}{2}}\frac{\partial_s r_\Lambda (q)}{1+r_\Lambda (y)}\left[ 1-\frac{\partial^2_z u_\Lambda (z)}{(1+r_\Lambda (y))y} +...\right] + (d-1)u_\Lambda (z) + \frac{3-d}{2} z\partial_z u_\Lambda (z)\; .   
\end{equation}
The condition of scale invariance of the flow equation requires that $r_\Lambda (y)$ can be written as a function of exclusively one dimensionless variable ($y$), i.e. no additional dependence on $\Lambda$ occurs \cite{Berges_02}. We may write $\partial_s r_\Lambda (q) = 2y\partial_y r (y)$. The first integral on the RHS of Eq.~(\ref{Expanded}) gives a physically irrelevant constant. The second one involves $\partial^2_z u_\Lambda (z)$ multiplied by a cutoff-dependent constant, which determines the capillary parameter $\omega$. We immediately observe that for $d=3$
\begin{equation}
\int dy y^{\frac{d-3}{2}}\frac{\partial_s r_\Lambda (q)}{(1+r_\Lambda (y))^2y} = 2\int_{r(y\to 0)}^{r(y\to\infty)} \frac{dr}{(1+r)^2}=-2\;,
\end{equation}
where we used $\lim_{y\to 0} r(y)=\infty$ and $\lim_{y\to \infty} r(y)=0$. This allows us to conclude that the scheme-dependence of the linear RG framework generically ceases as $d\to 3^-$.

The above fact and the observation that the terms neglected when
performing the linearization of the RG flow equation involve only higher powers
of the effective potential curvature, strongly support the reliability of the linear RG
approach in $d=3$. Cutoff dependence of the parameter $\omega$ would give rise to a
 scheme dependence of the critical exponents when these are computed within the linear RG
framework in $d<3$.

\section{Discussion} 
We end by discussing the assumptions of locality and lack of renormalization of the stiffness coefficient $\sigma$, which may seem controversial 
in view of more recent developments. In fact, the discrepancy between theoretical predictions based on the 
standard capillary-wave Hamiltonian Eq.~(\ref{Hamiltonian}) and numerical studies of wetting in the Ising model 
in $d=3$ has led to significant refinement of the former. The history of the development is reviewed in 
Ref.~\cite{Parry_09}. An attempt to resolve the abovementioned discrepancy led to the conclusion that the 
capillary-wave Hamiltonian Eq.~(\ref{Hamiltonian}) should be amended by a weak dependence of the stiffness 
coefficient $\sigma$ on $l$ \cite{Jin_93}. The RG studies of this model \cite{Jin_93_2, Boulter_98} revealed that the transition is driven 
weakly first order for a range of parameters. However, this seemingly improved capillary-wave model strongly contradicted 
other firmly accepted aspects of wetting theory \cite{Parry_09}. More recent developments suggest that a correct description of critical wetting 
phenomena requires a non-local interfacial model \cite{Parry_04, Parry_08}, which sheds light also on other puzzling features of wetting transitions 
including the wedge covariance property \cite{Parry_02, Greenall_04} and fulfillment of exact sum rules for the two-point correlation functions. 
The linear RG analysis of the proposed non-local model supports the predictions of the classical capillary-wave model regarding the 
non-universal critical behavior, simultaneously revealing the origin of the problem with the model retaining the dependence of 
$\sigma$ on $\tilde{l}$. Apart from subtleties, which however offer a
solution to the long-standing problem why the Ising model simulations do not resolve the critical 
region, the local capillary-wave Hamiltonian Eq.~(\ref{Hamiltonian}) turns out to serve as a good approximation to describe critical wetting 
behavior. 

The present work shows how the standard RG theory of wetting can be recovered
relying on few clear assumptions. The observation, that the key parameter of the RG theory ($\omega$)
is a scheme-dependent quantity makes the formal aspects of the approach
somewhat weaker, but does not deprive it of predictive power in the vicinity of the physical dimensionality $d=3$. Quite contrary, we showed that the ambiguities arise only below $d=3$ and the linear RG generically becomes perfectly robust against scheme variation in the limit $d\to 3^-$. 

\begin{acknowledgments}
I would like to thank Marek Napi\'{o}rkowski for useful discussions and a
number of comments on the manuscript. I also acknowledge a useful correspondence with Andrew O. Parry. 
\end{acknowledgments}

\end{document}